\documentclass[]{spie}  
\usepackage[]{graphicx}
\usepackage{amsmath}
\usepackage{amssymb}

\title{Direct measurement of the intra-pixel response function of the Kepler Space Telescope's CCDs} 

\author{Dmitry Vorobiev\supit{a}, Zoran Ninkov\supit{a}, Douglas Caldwell\supit{b}, and Stefan Mochnacki\supit{c}
\skiplinehalf
\supit{a}Center for Imaging Science, Rochester Institute of Technology, Rochester, NY, 14623, USA; \\
\supit{b}NASA Ames Research Center, Moffett Blvd, Mountain View, CA 94035, USA; \\
\supit{c}Department of Astronomy and Astrophysics, University of Toronto, Toronto, ON, Canada
}

\authorinfo{Further author information: (Send correspondence to Dmitry Vorobiev)\\ Dmitry Vorobiev: E-mail: vorobiev@cis.rit.edu}

\begin{document} 
  \maketitle 

\begin{abstract}
Space missions designed for high precision photometric monitoring of stars often under-sample the point-spread function, with much of the light landing within a single pixel. Missions like MOST, \textit{Kepler}, BRITE, and TESS, do this to avoid uncertainties due to pixel-to-pixel response nonuniformity. This approach has worked remarkably well. However, individual pixels also exhibit response nonuniformity. Typically, pixels are most sensitive near their centers and less sensitive near the edges, with a difference in response of as much as 50\%. The exact shape of this fall-off, and its dependence on the wavelength of light, is the intra-pixel response function (IPRF). A direct measurement of the IPRF can be used to improve the photometric uncertainties, leading to improved photometry and astrometry of under-sampled systems. Using the spot-scan technique, we measured the IPRF of a flight spare e2v CCD90 imaging sensor, which is used in the \textit{Kepler} focal plane. Our spot scanner generates spots with a full-width at half-maximum of $\lesssim$5 microns across the range of 400 nm - 900 nm. We find that Kepler's CCD shows similar IPRF behavior to other back-illuminated devices, with a decrease in responsivity near the edges of a pixel by $\sim$50\%. The IPRF also depends on wavelength, exhibiting a large amount of diffusion at shorter wavelengths and becoming much more defined by the gate structure in the near-IR. This method can also be used to measure the IPRF of the CCDs used for TESS, which borrows much from the \textit{Kepler} mission.
\end{abstract}



\section{INTRODUCTION}
\label{sec:intro}  
The \textit{Kepler} mission was designed to obtain very precise photometry over a wide field to detect extrasolar planets, especially Earth-like ones, as they transit their parent stars\cite{Borucki2010}. It has been spectacularly successful, finding 2,245 planet candidates, 2165 eclipsing binary stars and 2,342 confirmed exoplanets (as of April 11, 2018). Data collection for the original \textit{Kepler} mission ended in May 2013 and efforts are under way to develop a high quality public archive. In June 2014, the K2 mission became operational. 

\begin{figure}
    \centering
    \includegraphics[width=\textwidth]{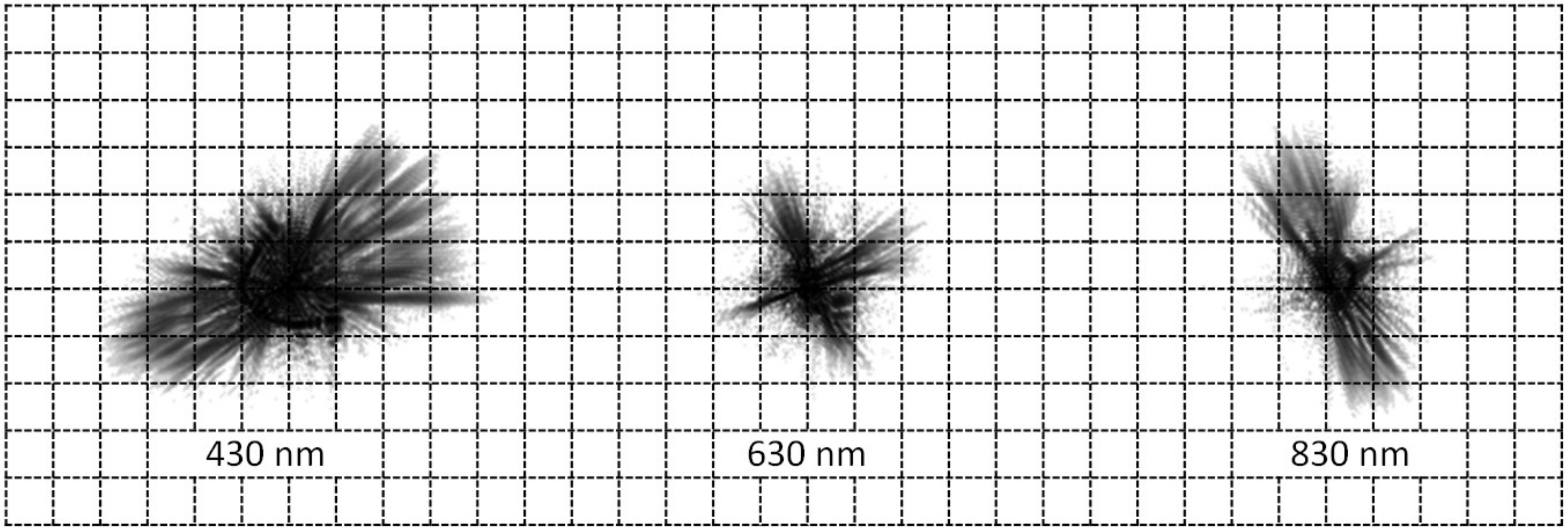}
    \caption{Synthetic monochromatic PSFs generated by \textit{Kepler}'s optical model show Chromatic Aberration in the PSF and the sampling of the images by the CCDs pixel grid. Each of the sub-sampled images is shown on an area of 11x11 pixels; the intensity is inverted (dark is higher intensity) and histogram-equalization scaled. The central pixel can contain up to 50\% of the total intensity. Adapted from Figure 14, \textit{Kepler} Instrument Handbook}
    \label{fig:kepler_psf}
\end{figure}

As in many space-based observatories, especially wide-field ones, stellar images recorded by \textit{Kepler}'s CCD detector matrix are under-sampled; i.e. the images formed on the CCD surface contain spatial frequencies higher than the Nyquist sampling frequency, which is set by the spacing/size of the CCD pixels. In \textit{Kepler}, the ``plate scale'' is 3.98 arcseconds per pixel. While  stellar images (point spread function) have a 95\% enclosed energy diameter of 6.4 pixels, the central pixel can contain up to 50\% of the energy (i.e. Brightest Pixel Flux Fraction $\leq$ 0.5), which indicates a full width at half maximum of less than 2 pixels, the Shannon criterion for full sampling. The point spread function of the telescope can have spikes and other components of high spatial frequency (Figure \ref{fig:kepler_psf}), which exacerbate problems that arise from the under-sampling of stellar images.

\textit{Kepler} photometry has until now been based on aperture photometry using a so-called “pixel response function”, which is really a super-sampled point spread function of the entire system\cite{Bryson2010}, including optical effects of the telescope, pointing jitter and the underlying pixel response structure. This is now being improved in anticipation of photometry by fitting stellar profiles (``PRF fitting'') rather than photometry by simple integration of signal over an aperture\cite{Morris2014}. 

The outstanding precision of \textit{Kepler} aperture photometry, of about 10 parts in a million over 6 hours for 10th magnitude stars in its full original “K1” mission\cite{Christiansen2012} led to the anticipation that astrometry of 1 milliarcsecond accuracy might be possible, but in practice only about 4 milliarcsecond precision for a single measurement has been attained\cite{Monet2010}, with better precision being attained in certain cases. So far, centroid analysis of various kinds has been tried rather than PSF-fitting, but PRF fitting is now being implemented, although the astrometric benefits of this technique have not yet been published\cite{Morris2014}.

\subsection{K2 and Reduced Pointing Stability}

The unprecedented photometric precision achieved by \textit{Kepler} was largely made possible by the mitigation of many systematic errors having to do with the varying photometric sensitivity across the focal plane. These effects arise both from the optical design of the telescope (varying image quality across the large field of view) and from the pixel-to-pixel variations of the CCDs. \textit{Kepler} avoids both of these error sources by maintaining precise pointing as it orbits the sun.  The original \textit{Kepler} mission had a pointing stability (RMS jitter) of 0.063 pixels. The K2 mission has an increased RMS jitter of $\sim$0.5 pixels. This means that stars will drift across the pixels during and between exposures. Detailed knowledge of the IPRF will be critical to mitigating the photometric errors associated with this drift. 

\section{\textit{Kepler} Data Analysis and the Pixel Response Function}

\begin{figure}
    \centering
    \includegraphics[width=\textwidth]{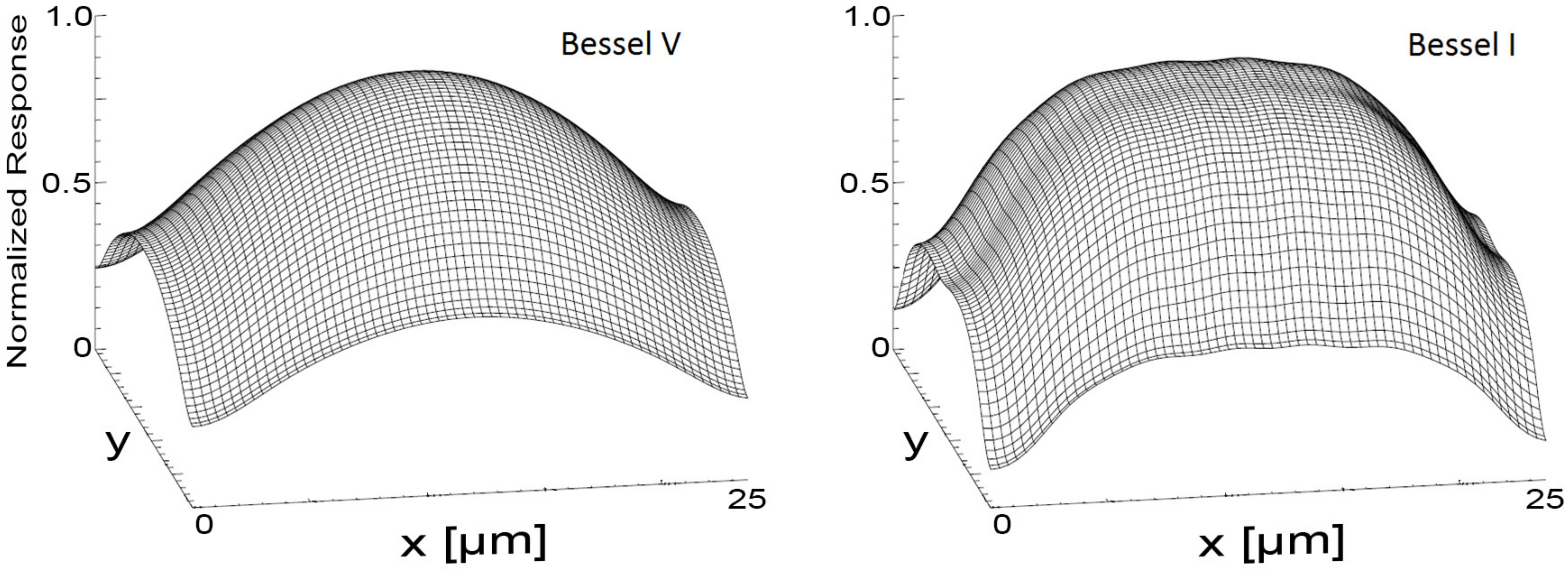}
    \caption{The IPRF of a SITe 502A backside illuminated CCD, measured in two broadband filters. In each case, the IPRF peaks near the physical center of the pixel and drops off rapidly towards the edges. However, the IPRF also depends significantly on the wavelength of light. Adapted from Piterman \& Ninkov 2002.}
    \label{fig:previous_iprf}
\end{figure}

\textit{Kepler} image analysis has hitherto followed the general prescription set by Lauer (1999)\cite{Lauer1999}, which requires the determination of a super-sampled PSF, which is the convolution of the optical PSF of the telescope system and of the spatial sensitivity of a single pixel (Eq. \ref{eq:prf}). We call the spatial variability in sensitivity of a pixel the ``intrapixel response function'' (IPRF). An ideal pixel has an IPRF that is spatially and temporally constant. In practice, this is never the case and intrapixel variation from 2\%-50\% has been measured for back-thinned CCDs\cite{Jorden1994,Piterman2002} (Figure \ref{fig:previous_iprf}). If left uncalibrated, this uncertainty in pixel response creates large errors in photometry and astrometry\cite{Bryson2010}. 

\subsection{The Intra-Pixel Response Function}

If one assumes that the intra-pixel sensitivity over a CCD is the same for each pixel (Jorden et al. 1994) and that it can be precisely determined, the intra-pixel response function can be separated from the optical PSF of the telescope.  (To avoid confusion, we will refer to the intra-pixel response function as the IPRF). Note that the IPRF can actually extend over several pixels, because the signal generated within one pixel can spill over to neighboring pixels; this can be caused by electron leakage and by the leakage of photons caused by a rapidly converging beam. 

Following Lauer (1999)\cite{Lauer1999}, we can write the super-sampled PSF or ``PRF'' as:
\begin{equation}
\label{eq:prf}
    PRF(x,y) = PSF(x,y)*IPRF(x,y)		 
\end{equation}

where $PSF$ is the ``optical'' PSF (including jitter),  $IPRF$ is the intra-pixel response function and $*$ the convolution operator. Spillage effects between pixels should be included in IPRF, which in fact spans a number of pixels, if necessary.

\section{Direct Measurement of the IPRF}
If the IPRF is well-characterized, the PSF can be precisely determined for each star in a field using knowledge of the optics and further refined with actual observations. A time-consuming raster of on-orbit observations to measure the ``PRF'' would no longer be required and the under-sampling effects could be corrected systematically, assuming that $IPRF(x, y)$ is the same for all pixels. 

The IPRF has been measured for frontside-illuminated\cite{Jorden1994, Kavaldjiev1998} and backside-illuminated devices\cite{Jorden1994, Piterman2002}. Typically, a single spot with a diameter which is smaller than the size of a pixel ($\sim$ 1 - 5 $\mu m$) is rastered across a set of several pixels to produce an X-Y grid with spacing much smaller than the size of a pixel. The measurements are then interpolated to estimate the IPRF (Figure \ref{fig:previous_iprf}).

\subsection{Measurement Setup}
\begin{figure}
    \centering
    \includegraphics[width=\textwidth]{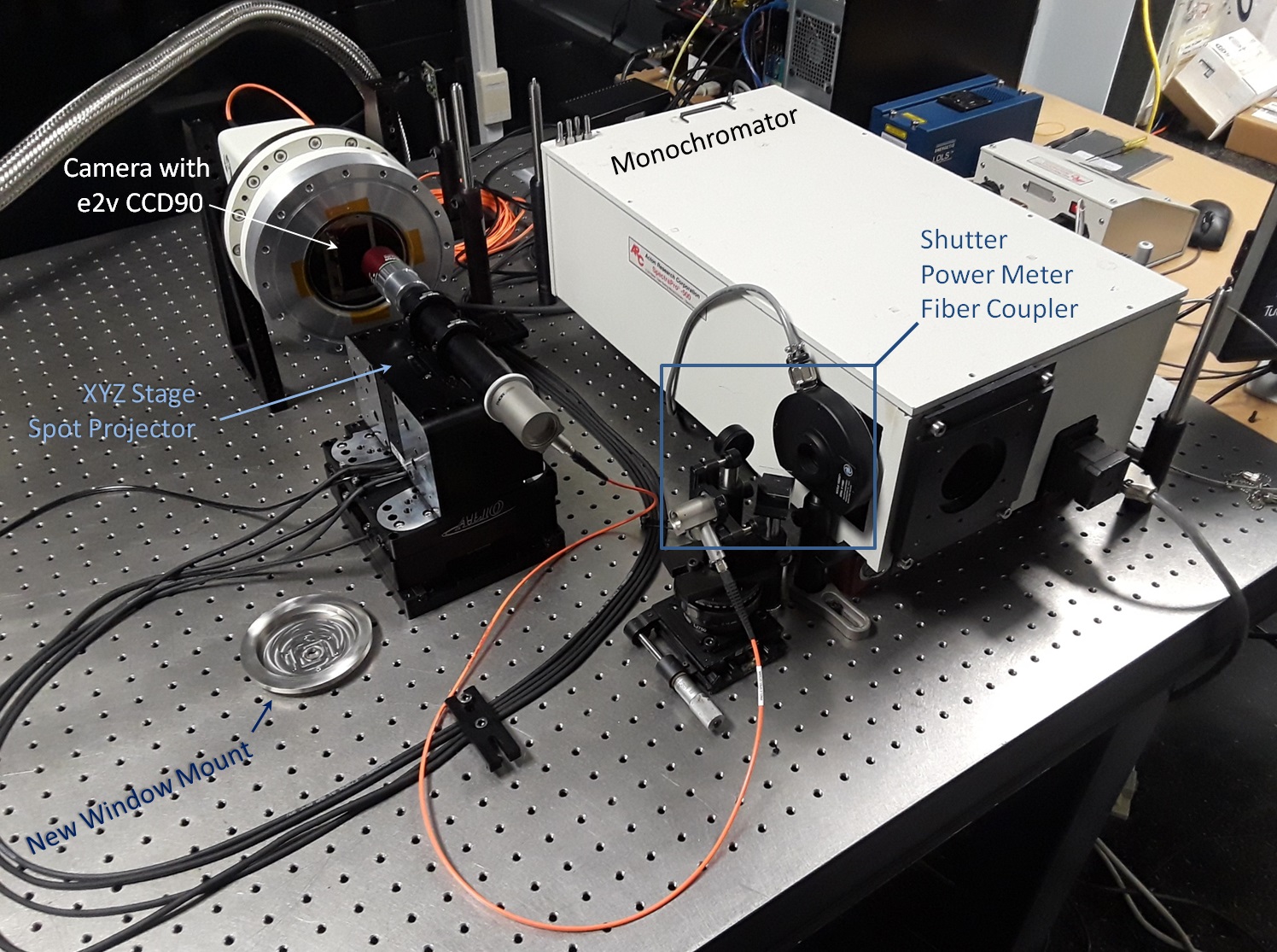}
    \caption{A new apparatus to measure the intra-pixel response function of a range of imaging arrays was built in RIT's Laboratory for Advanced Instrumentation Research. This system is able to produce small spots of light, across a wide range of wavelengths.}
    \label{fig:spotscanner}
\end{figure}

A technique for the direct measurement of intra-pixel sensitivity variations has been developed and used at Rochester Institute of Technology for a number of years \cite{Kavaldjiev1998, Piterman2002}. Building on our previous experience, we built a new high performance measurement apparatus dedicated to the measurement of the IPRF (Figure \ref{fig:spotscanner}). This system is capable of producing spots with full width at half maximum (FWHM) of 4 microns or less (see Section \ref{sec:spotCharacterization}), at spacing intervals as small as 0.1 $\pm$ 0.02 $\mu m$. This allows us to sample \textit{Kepler}'s 27 $\mu m$ pixels with a wide range of sampling grids - from coarse (10 $\times$ 10) to extremely fine (270 $\times$ 270). A brief description of the major components is given below. 

\subsubsection{Light Source}
In this setup, we use an Energetiq EQ-99XFC Laser Driven Light Source, which produces a stable broadband spectrum. The source stability during a measurement is further monitored by a Thorlabs PM-100D power meter. Over the course of a typical measurement, the source output varies by $\sim$1.5\% rms on the scale of seconds; however, over the course of hours the instrument is stable to better than 0.1\%.

\subsubsection{Wavelength Selection}
\begin{figure}
    \centering
    \includegraphics[width=\textwidth]{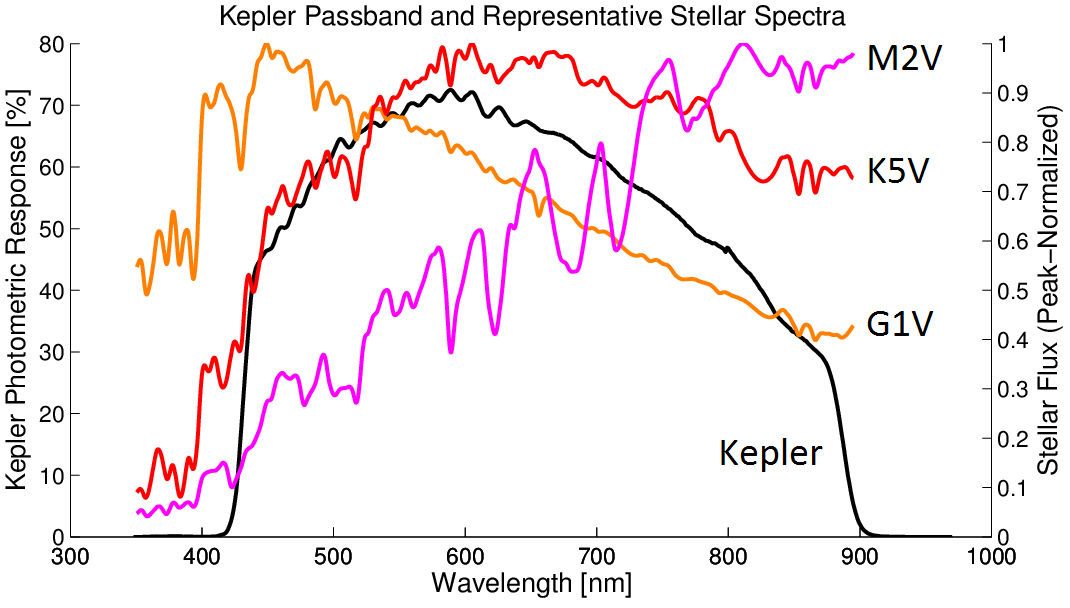}
    \caption{\textit{Kepler} observes stars of different colors through one passband. Both the optical PSF of the telescope and the IPRF change with wavelength. Therefore, calibrating the spectral dependence of the IPRF is extremely beneficial. Our measurement apparatus operates across the entire \textit{Kepler} passband.}
    \label{fig:stellarSpectra}
\end{figure}

The broadband light source is filtered using an Acton 0.5 meter monochromator, to produce light with a spectral bandwidth of $\Delta \lambda\approx$15 nm. This allows us to characterize the dependence of the IPRF on wavelength. The wavelength dependence is especially important for the \textit{Kepler} photometer, due to its broad passband and the large variation in the spectral distribution of stars in the visible range (Figure \ref{fig:stellarSpectra}). Measurements in the spectral range longward of 700 nm are made by inserting a Bessel I filter at the entrance slit of the monochromator, to block the higher order UV and optical light.  

\subsubsection{Spot Projection} 
A wavelength-dependent characterization of \textit{Kepler}'s IPRF requires the ability to produce small-diameter spots across a wide range of wavelengths. For this purpose, we use a 20X Mitutoyo Plan Apo NIR infinity corrected objective. These long working distance objectives are well suited for illumination of sensors that must be housed in vacuum camera heads (such as the e2v CCD90). However, these fast objectives must be used with sufficiently thin windows, so as to not induce significant spherical aberration (see Section \ref{fig:windowZoom}). 

\subsubsection{Spot Translation}
The spot projector is fiber-fed and mounted on XYZ translation stage from Alio Industries. To be able to use a variety of cameras/dewars with the \textit{Kepler} CCD, we designed the system to translate the spot projector (rather than the sensor). The optical system is rigid to prevent sagging. The XYZ translation mechanism is capable of step sizes as small as 2 nm, with a bidirectional repeatability of 20 nm. This is an order of magnitude better than our minimum step size (250 nm). We chose this system for its excellent repeatability and thermal stability (the scale for the encoder is made from ZERODUR glass). To reduce vibrational jitter, this apparatus is setup in a basement laboratory whose foundation is independent of the rest of the building.

\subsubsection{CCD Camera}
\label{sec:setupCCDCamera} 
The e2v CCD90 (SN 208) was acquired through NASA Ames Research Center, from Ball Aerospace Corporation. The sensor was installed in a Spectral Instruments 800 Series camera. This camera allows us to cool the sensor to -35$^{\circ}$ C, which is significantly warmer than \textit{Kepler}'s science operating temperature of -85$^{\circ}$ C. The temperature we reach in the lab (-35$^{\circ}$ C) reduces the dark current to a manageable level; however, the absorption length in silicon depends on temperature, as well as wavelength. Therefore, a wavelength-temperature correction is needed to relate our measurements to \textit{Kepler}'s on-orbit performance (see Section \ref{sec:futureWork} for a further discussion).

\begin{figure}
    \centering
    \includegraphics[width=\textwidth]{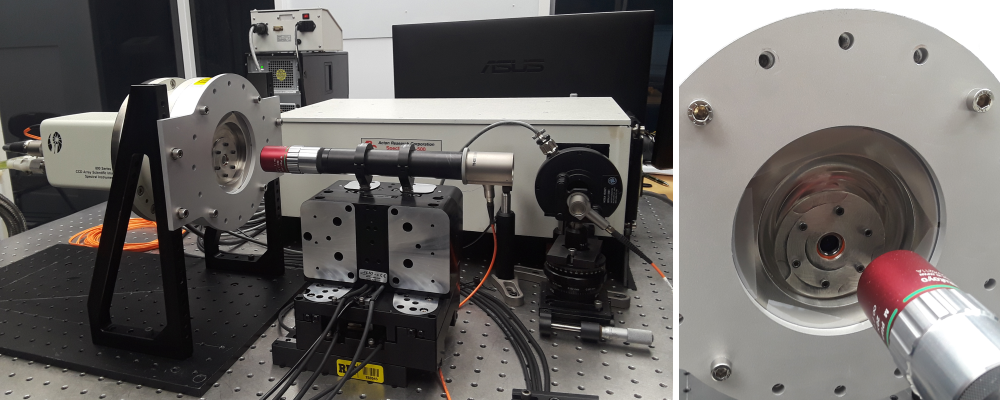}
    \caption{The camera's normal 4 inch window was replaced with a much smaller, thinner fused silica window, which is mounted in a stainless steel frame.}
    \label{fig:windowZoom}
\end{figure}

The CCD camera's original 4 inch diameter window was replaced with a steel frame and a much smaller fused silica window. The new window is 500 $\mu m$ thick, with a diameter of 10 mm (Figure \ref{fig:windowZoom}). This significantly reduces the areas on the sensor which are accessible for measurement. However, the remaining area is sufficiently to measure thousands of pixels. Furthermore, the window was placed off-axis, so that different regions of the sensor can be probed by rotating the mount. 

\subsection{Spot Characterization}
\label{sec:spotCharacterization}
The spot size was measured at 550 nm using an IDS uEye UI-1492LE CMOS camera. This camera uses a sensor with 1.67 $\mu m$ pixels. The spot was approximately centered on a single pixel, and a set of multiple exposures ($\sim$50) were acquired. The individual exposures were combined using a median filter to reduce read noise and dark noise. To characterize the spot diameter, a 2D Gaussian function was fit to the data. The $\sigma_x$ and $\sigma_y$ parameters of the best fitting Gaussian were used to estimate the full width at half maximum (FWHM) of the incident spot, using the relation below,

\begin{align*}
    \text{FWHM} = 2 \sqrt{2 \text{ln}2} \sigma.
\end{align*}

\begin{figure}
    \centering
    \includegraphics[width=\textwidth]{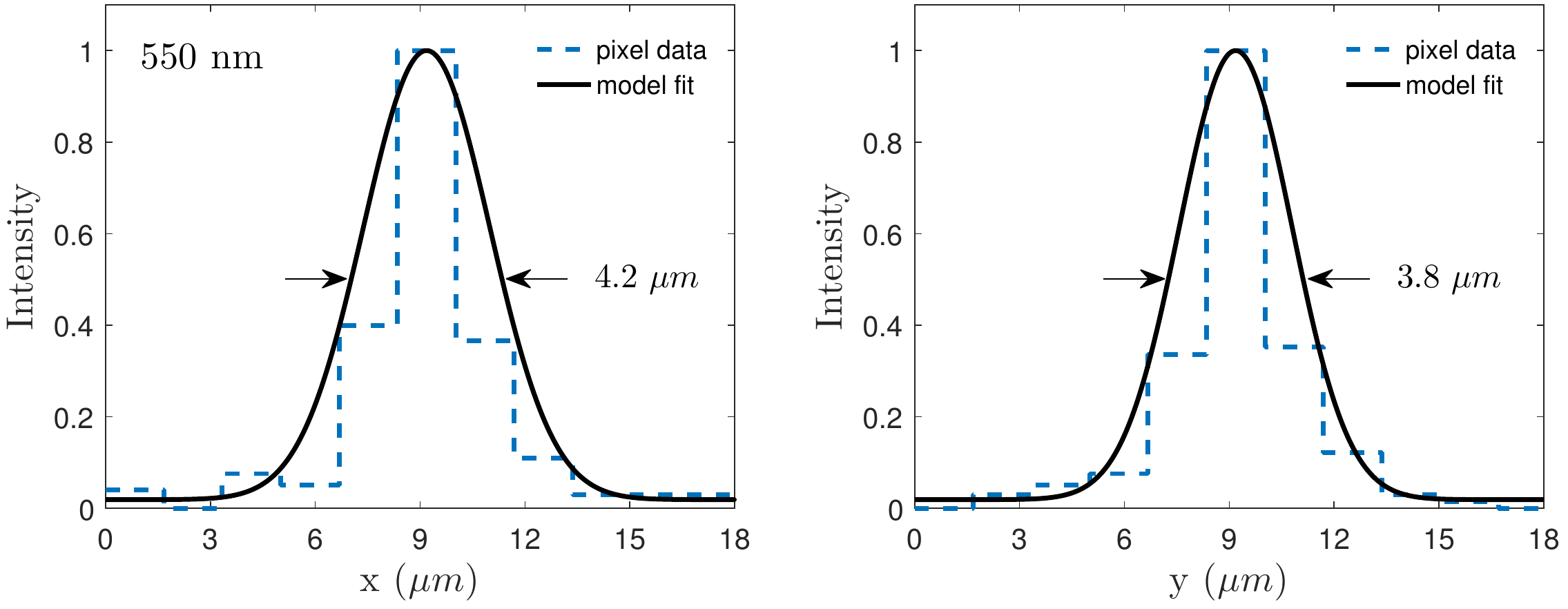}
    \caption{The normalized profile of the spot size obtained at 550 nm, in the \^{x} (Left) and \^{y} directions (Right). Each pixel spans 1.67 $\mu m$. The spot appears symmetric, within the measurement uncertainty.}
    \label{fig:spotSize_noWindow}
\end{figure}

The resulting pixel data and the best fit model are shown in Figure \ref{fig:spotSize_noWindow}. The FWHM estimated for the spot in the \^{x} and \^{y} directions are 4.2 $\pm$ 0.2 $\mu m$ and 3.8 $\pm$ 0.2 $\mu m$, respectively. Considering the uncertainties of the fit parameters and spot positioning, the spot appears symmetric. We expect that this estimate is an upper bound, because the 1.67 $\mu m$ pixels are still quite large compared to the spot size we're trying to estimate. 

\subsubsection{Effects of the Camera Window on Spot Size}
\begin{figure}
    \centering
    \includegraphics[width=\textwidth]{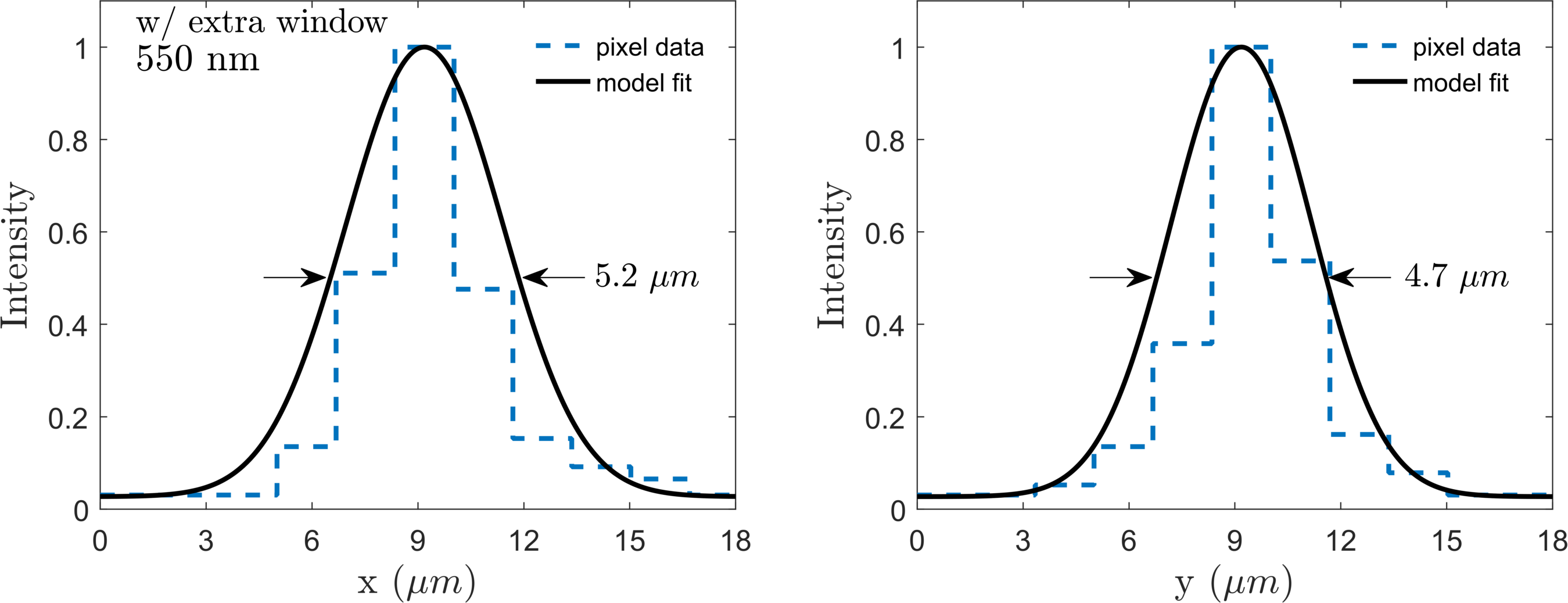}
    \caption{The normalized profile of the spot size obtained at 550 nm, in the \^{x} (Left) and \^{y} directions (Right), with a 500 $\mu m$ thick window introduced into the beam. Each pixel spans 1.67 $\mu m$.}
    \label{fig:spotProfile_withWindow}
\end{figure}

The e2v CCD90 at the focus of this work is housed in a vacuum camera head, with a 500 $\mu m$ thick fused silica window. Some degradation of the optical performance of the Mitutoyo objective is expected, due to the spherical aberration which arises when a window is introduced into the converging beam. To estimate the magnitude of this effect, the measurement described in the previous section (Sec. \ref{sec:spotCharacterization}) was repeated, with an identical 500 $\mu m$ window in the beam. The system was refocused (to account for the change in focal distance). The resulting spot measurement is shown in Figure \ref{fig:spotProfile_withWindow}. 

The spot measured in this configuration is $\sim$ 5 $\mu m$ FWHM. Note that the uEye UI-1492LE camera already has a 400 $\mu m$ thick window, close to the sensor. Therefore, we do not have a direct measurement of the spot size, the way it appears on the \textit{Kepler} CCD. However, this measurement with an extra window can be considered an upper bound to the spot size. We are currently developing more accurate methods to characterize these small spots.

\subsection{Measurement Results}
\begin{figure}
    \centering
    \includegraphics[width=\textwidth]{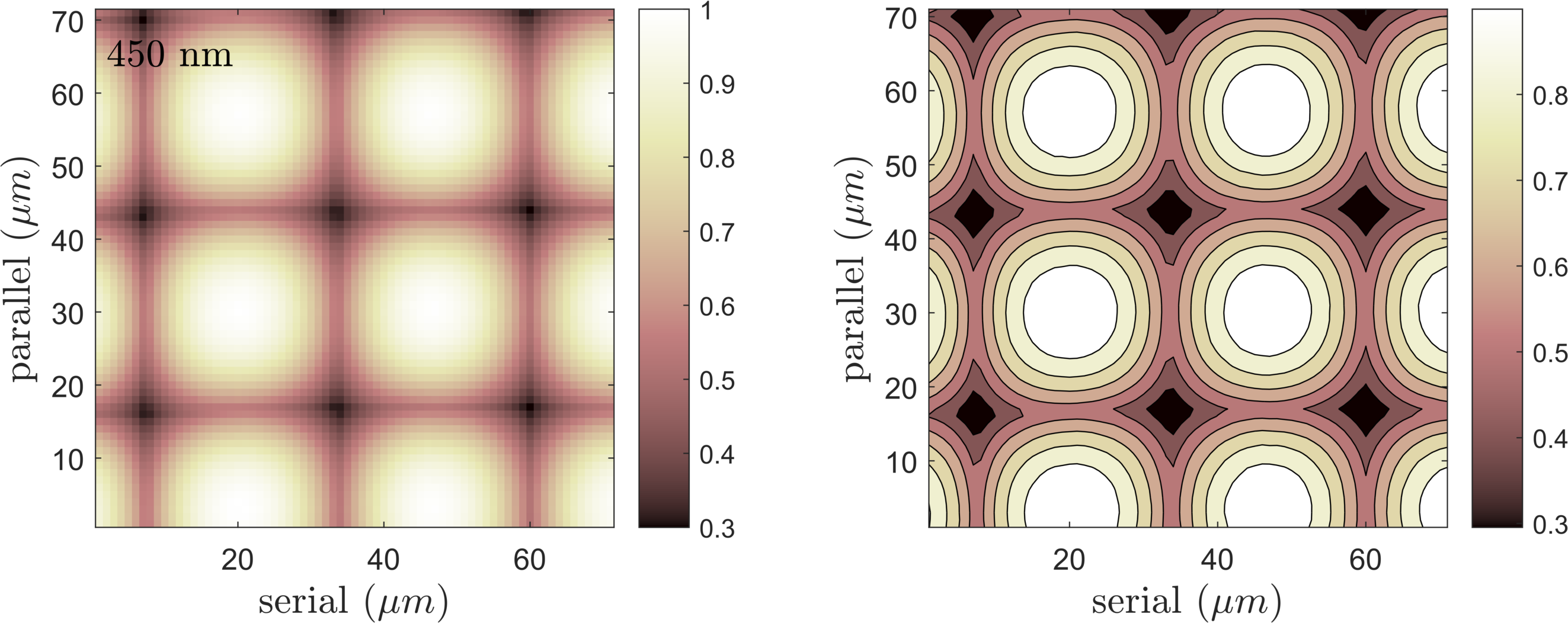}
    \caption{The normalized intra-pixel response of \textit{Kepler}'s CCD at 450 nm, as measured by a spot scanner at RIT. The response is maximal the central third of a pixel, and rapidly decreases towards the pixel edges.}
    \label{fig:450nm_fullScan}
\end{figure}

Our initial scans consisted of a 70 $\mu m$ $\times$ 70 $\mu m$ region of interest, with steps of 1 $\mu m$ in each direction. The measurements were performed at 450 nm, 600 nm, and 800 nm (with a passband of $\Delta \lambda \sim 15$ nm). The entire scan region at 450 nm is shown in Figure \ref{fig:450nm_fullScan}. To determine the IPRF, we normalize the intensity measured by the brightest pixel at each position of the spot scanner. This relative measurement must be combined with the quantum efficiency of the CCD at the appropriate wavelength to obtain the absolute IPRF. In the following sections, we consider the IPRF at our three wavelengths of interest.   

\subsubsection{Response at 450 nm}
\begin{figure}
    \centering
    \includegraphics[width=\textwidth]{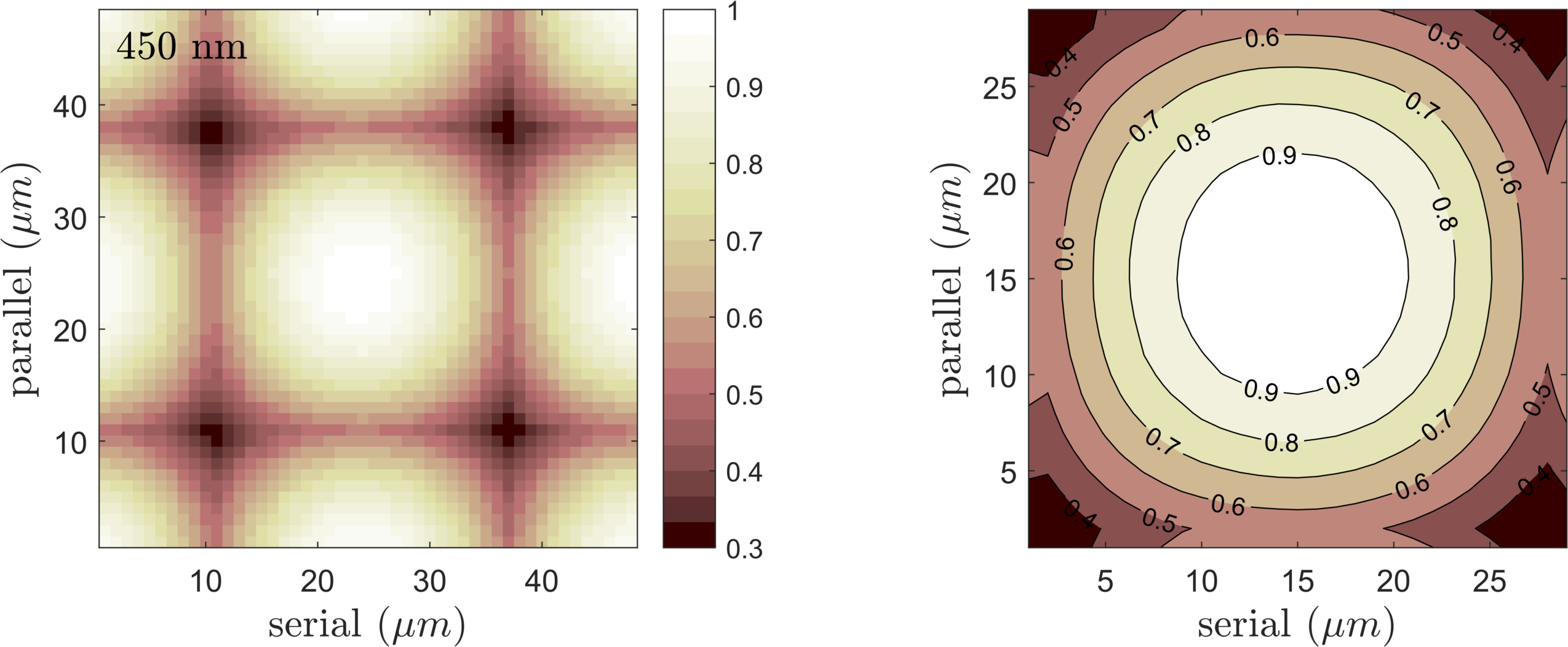}
    \caption{The normalized intra-pixel response of \textit{Kepler}'s CCD at 450 nm, as measured by a spot scanner at RIT. The same data is shown using an intensity map (Left) and a contour plot (Right). The response is maximal in the central third of a pixel, and rapidly decreases towards the pixel edges.}
    \label{fig:450nm_zoom}
\end{figure}

\begin{figure}
    \centering
    \includegraphics[width=\textwidth]{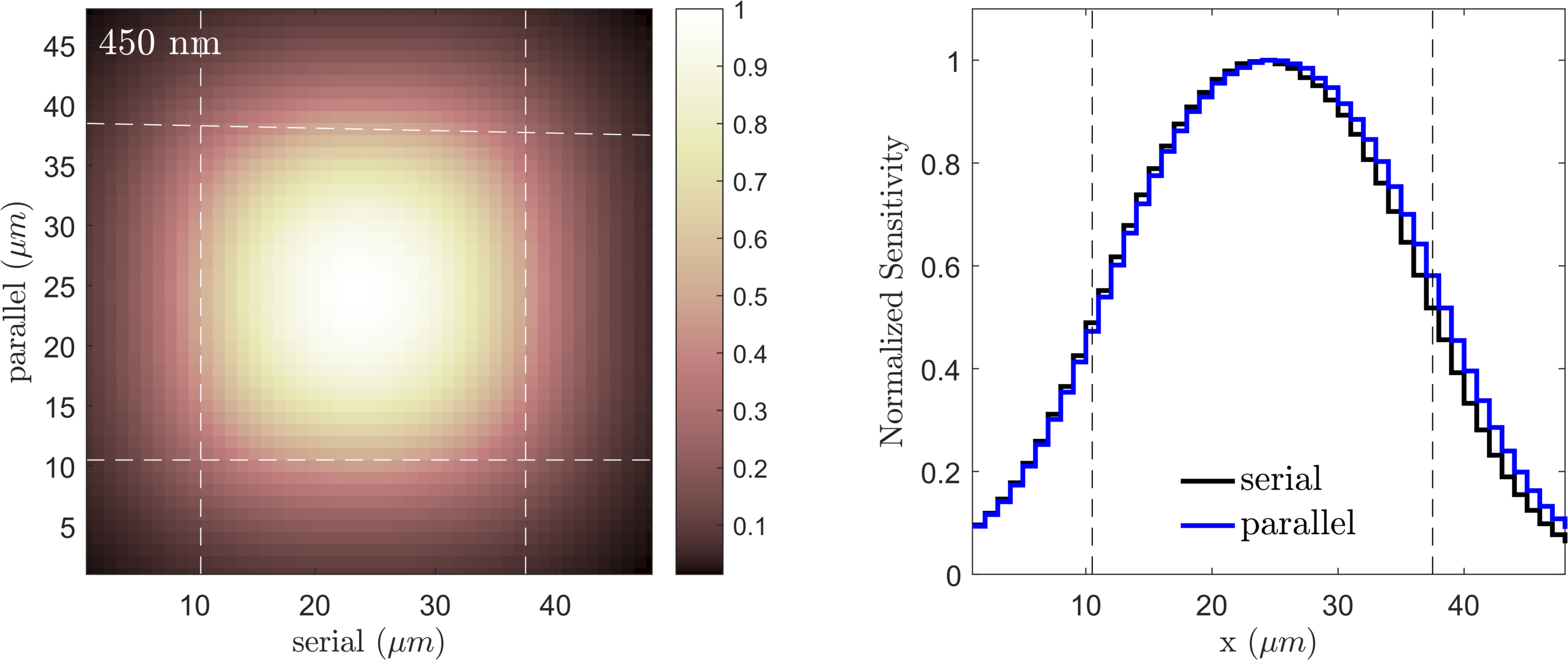}
    \caption{The normalized response of a single pixel to 450 nm light, as a spot is scanned across. The axes show the position of the spot center. The left panel shows an intensity map of the 2D distribution, while the right panel shows a profile across the middle of the pixel, in the serial and parallel directions.}
    \label{fig:450nm_profile}
\end{figure}

First, we look at the pixel response at 450 nm, which is towards the blue end of the \textit{Kepler} passband (Figure \ref{fig:stellarSpectra}). A zoomed-in view of the IPRF is shown in Figure \ref{fig:450nm_zoom}. The response is similar to other back-illuminated devices, showing a considerable amount of diffusion and round contours. The peak response decreases by $\sim$70\% at the corners of each pixel. Another way to characterize this diffusion is to measure the response of a single pixel, as the spot is scanned across. The resulting map is shown in Figure \ref{fig:450nm_profile}. 

\subsubsection{Response at 600 nm}
\begin{figure}
    \centering
    \includegraphics[width=\textwidth]{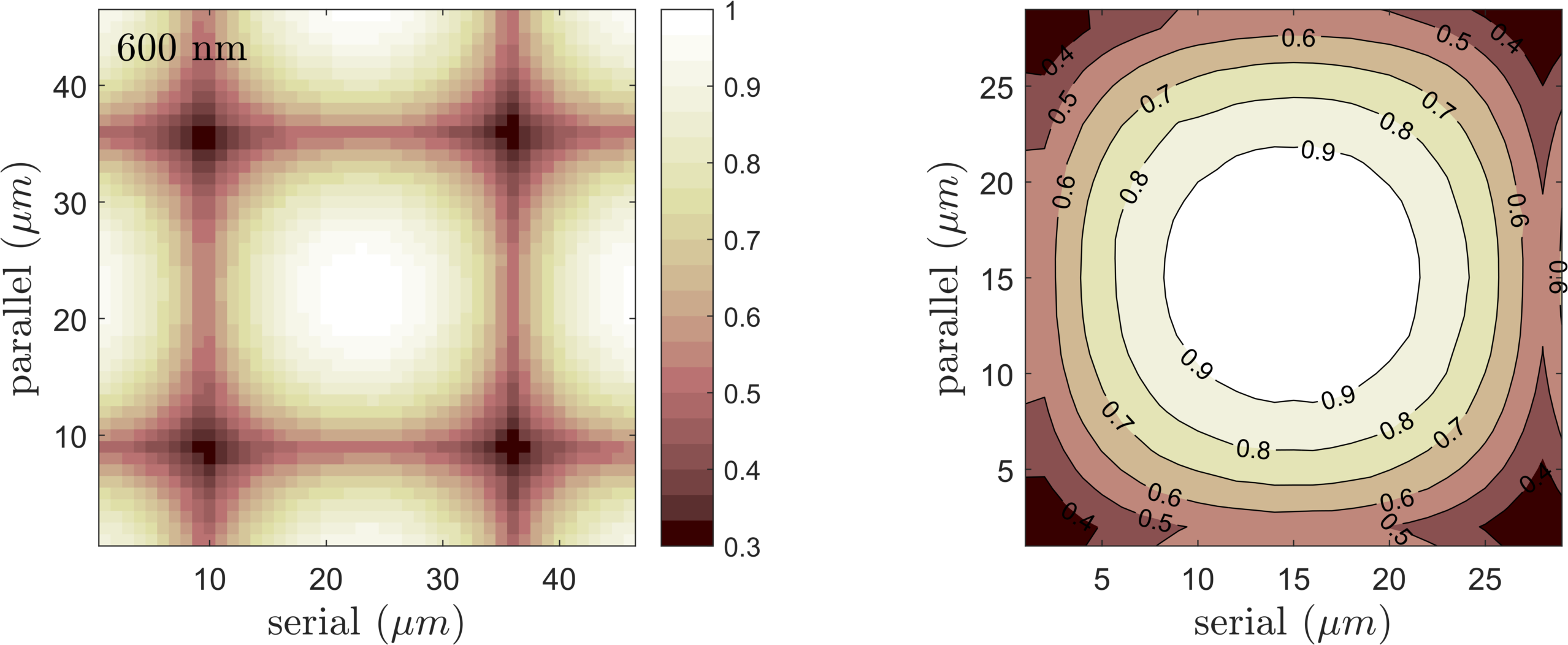}
    \caption{The normalized intra-pixel response of \textit{Kepler}'s CCD at 600 nm, as measured by a spot scanner at RIT. The same data is shown using an intensity map (Left) and a contour plot (Right). The response is maximal in the central third of a pixel, and rapidly decreases towards the pixel edges.}
    \label{fig:600nm_zoom}
\end{figure}

\begin{figure}
    \centering
    \includegraphics[width=\textwidth]{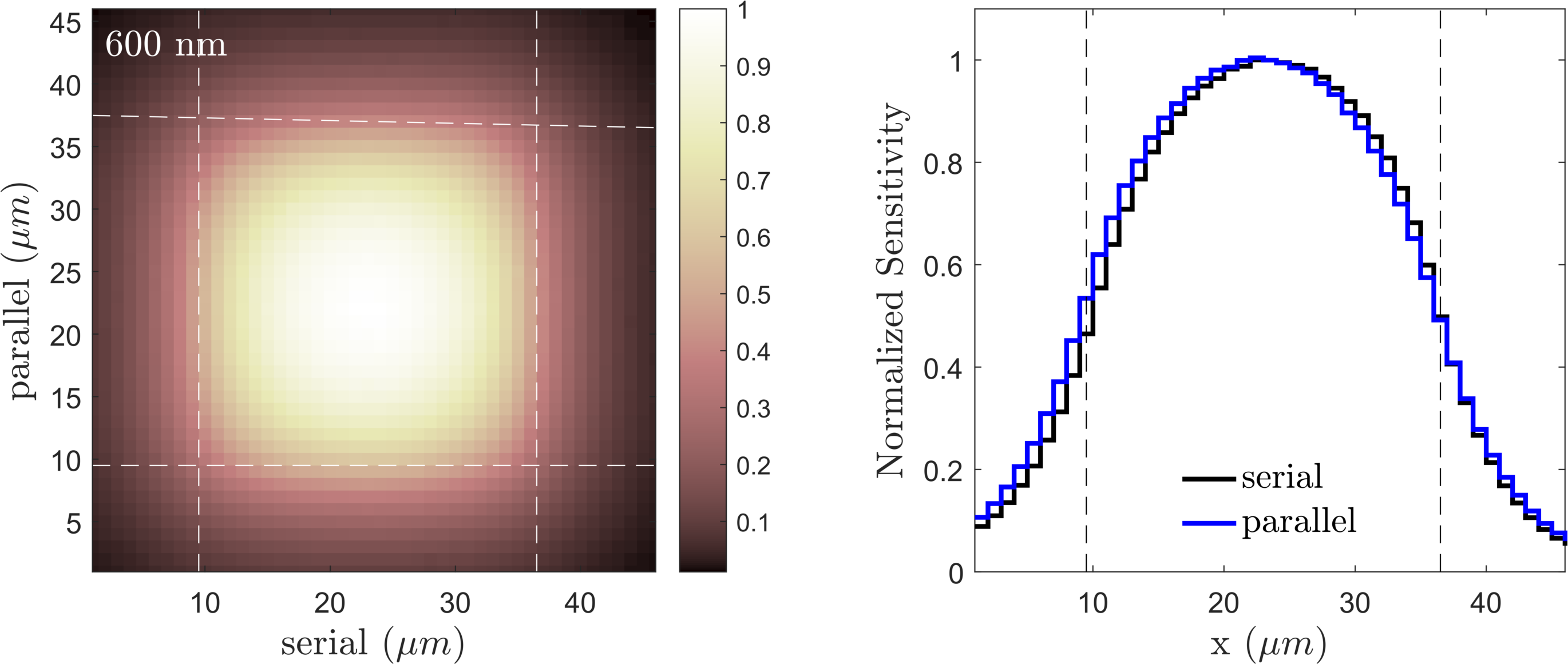}
    \caption{The normalized response of a single pixel to 600 nm light, as a spot is scanned across. The axes show the position of the spot center. The left panel shows an intensity map of the 2D distribution, while the right panel shows a profile across the middle of the pixel, in the serial and parallel directions.}
    \label{fig:600nm_profile}
\end{figure}

Next, we present the results of our spot scan at 600 nm ($\Delta \lambda \approx$ 15 nm). The intensity maps of the intra-pixel response function are shown in Figure \ref{fig:600nm_zoom} and Figure \ref{fig:600nm_profile}. Overall, the results are similar to the response at 450 nm, though the potential well of a single appears slightly more well-defined.

\subsubsection{Response at 800 nm}
\begin{figure}
    \centering
    \includegraphics[width=\textwidth]{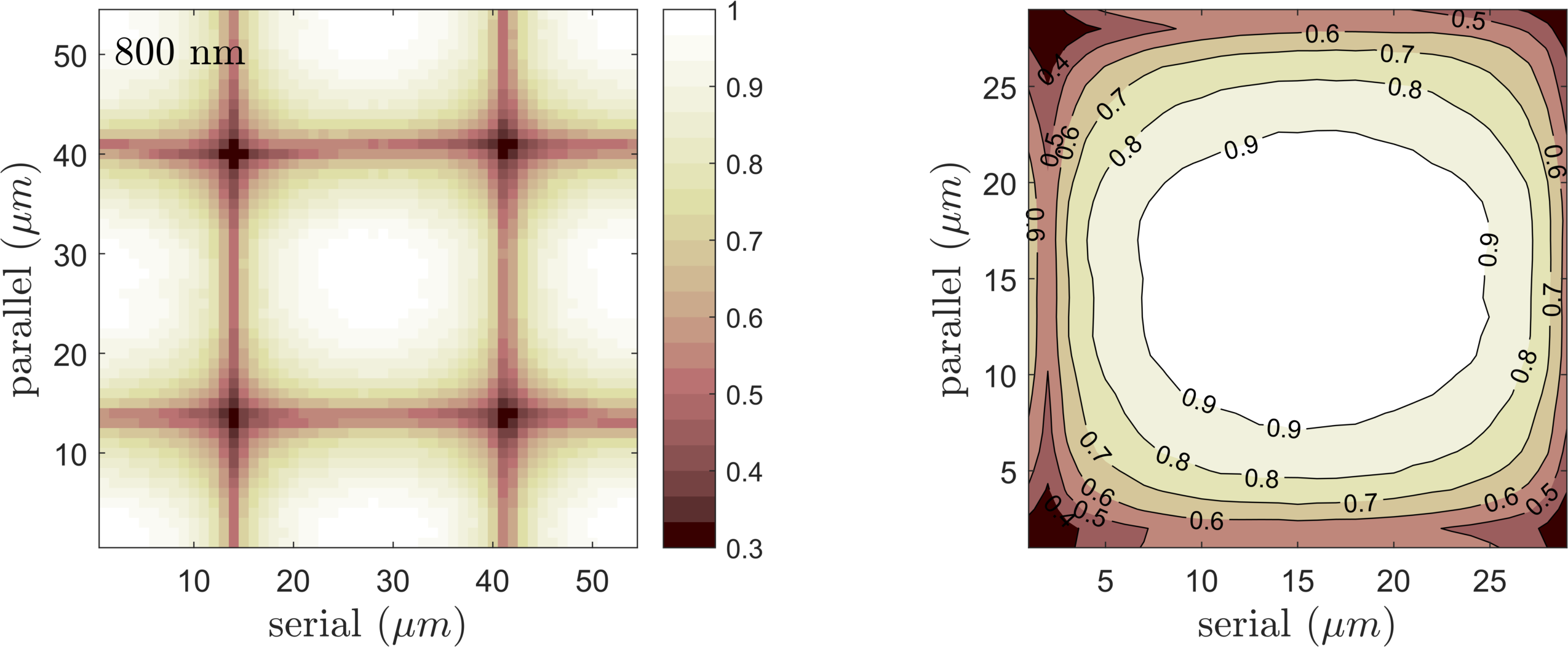}
    \caption{The normalized intra-pixel response of \textit{Kepler}'s CCD at 800 nm, as measured by a spot scanner at RIT. The same data is shown using an intensity map (Left) and a contour plot (Right). The response is maximal in the central third of a pixel, and rapidly decreases towards the pixel edges.}
    \label{fig:800nm_zoom}
\end{figure}

\begin{figure}
    \centering
    \includegraphics[width=\textwidth]{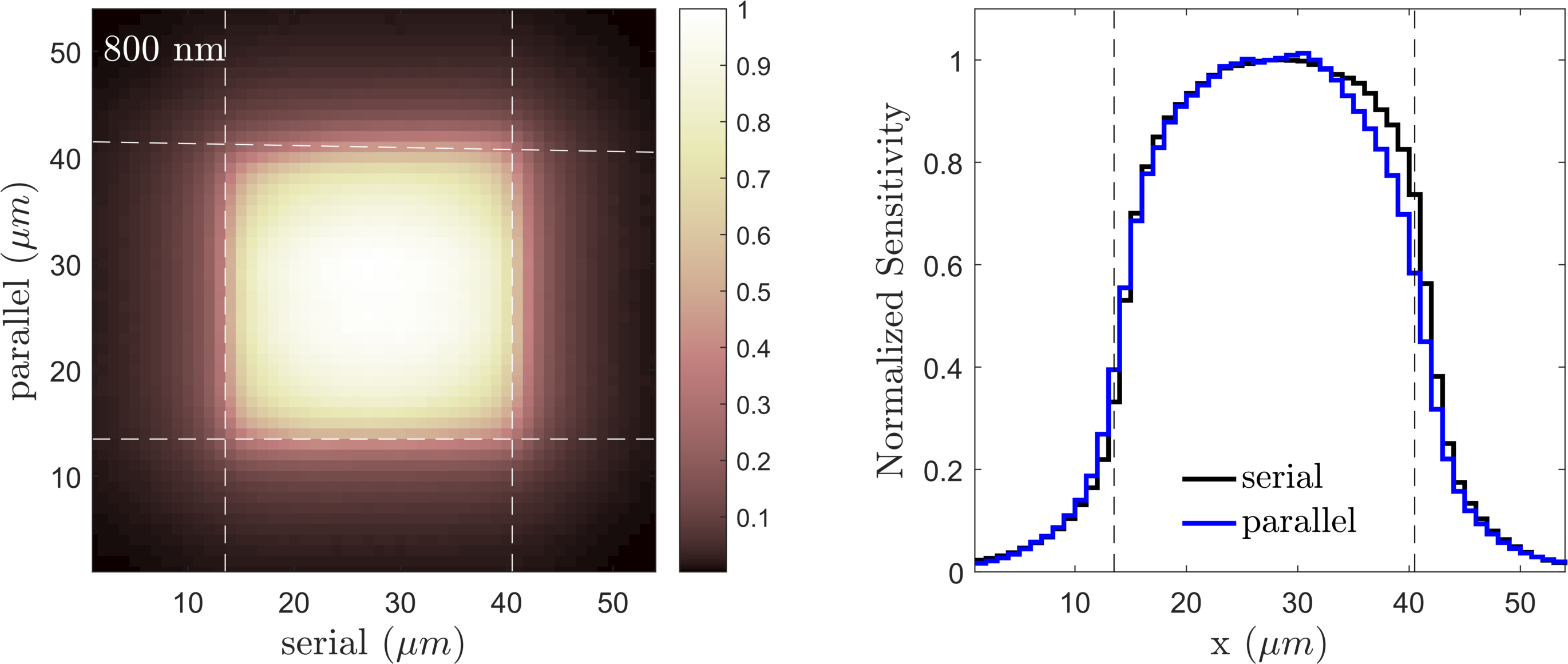}
    \caption{The normalized response of a single pixel to 800 nm light, as a spot is scanned across. The axes show the position of the spot center. The left panel shows an intensity map of the 2D distribution, while the right panel shows a profile across the middle of the pixel, in the serial and parallel directions.}
    \label{fig:800nm_profile}
\end{figure}

Lastly, we show the IPRF we estimated at 800 nm in Figure \ref{fig:800nm_zoom} and Figure \ref{fig:800nm_profile}. The IPRF at longer wavelengths is noticeably different than at shorter wavelengths. The IPRF is much better defined, as can be seen by the shape of the contours in Figure \ref{fig:800nm_zoom} (Right panel) and the profile shown in Figure \ref{fig:800nm_profile} (Right panel). 

\clearpage
\section{Discussion of Results}
We have presented the design and the first results of our effort to directly characterize the intra-pixel response function of the e2v CCD90 sensor, across the passband of the \textit{Kepler} photometer. We have shown that our spot projector is able to generate spots of FWHM $\sim$ 4 - 5 $\mu m$, at 550 nm. Furthermore, the source and the mechanical translation system appear quite stable, as can be inferred from the very consistent response measured over the course of several hours (Figure \ref{fig:450nm_fullScan}).

\subsection{\textit{Kepler'}s IPRF}
The IPRF we measured for \textit{Kepler}'s CCD appears similar to the IPRF measured for other back-illuminated CCDs\cite{Jorden1994, Piterman2002} (Figure \ref{fig:previous_iprf}). Two main trends can be seen in these initial data: 

\begin{enumerate}
    \item The IPRF varies significantly between the center and edges of a pixel. The sensitivity at the edge of a pixel is $\sim$50\% lower than at the pixel's center, and as much as 70\% lower near the pixel corner.
    \item The IPRF is a function of wavelength, as well as position. The IPRF in the near-IR is noticeably different than in the visible range. 
\end{enumerate}

The spot scanning technique also allows for a systematic analysis of the diffusion process in imaging arrays. Examples of this measurement are shown in the left panels of Figure \ref{fig:450nm_profile}, Figure \ref{fig:600nm_profile}, and \ref{fig:800nm_profile}. In these maps, the effects of IPRF and charge diffusion are not disentangled from the shape of the illuminating spot. However, the strong wavelength dependence of these profiles suggests that the effects of the spot shape are secondary to the diffusion process. 

\section{Future Work}
\label{sec:futureWork}
The goal of this work is to directly measure the IPRF of \textit{Kepler}'s CCD, as a function of wavelength, to improve the photometric and photometric precision of \textit{Kepler} and \textit{K2} science. This requires high resolution IPRF maps at many wavelengths across the \textit{Kepler} passband. Some further development of our system remains, before these maps can be incorporated into the data reduction pipelines. 

First, some aspects of the measurement setup need further investigation:
\begin{itemize}
    \item The spot size and shape should be re-measured at each wavelength of interest; ideally, with a small-pixel camera and a window configured identically to the measurement setup.
    \item The repeatability of the measurements should be verified quantitatively by comparing several scans with identical parameters. 
\end{itemize}

\begin{figure}
    \centering
    \includegraphics[width=0.7\textwidth]{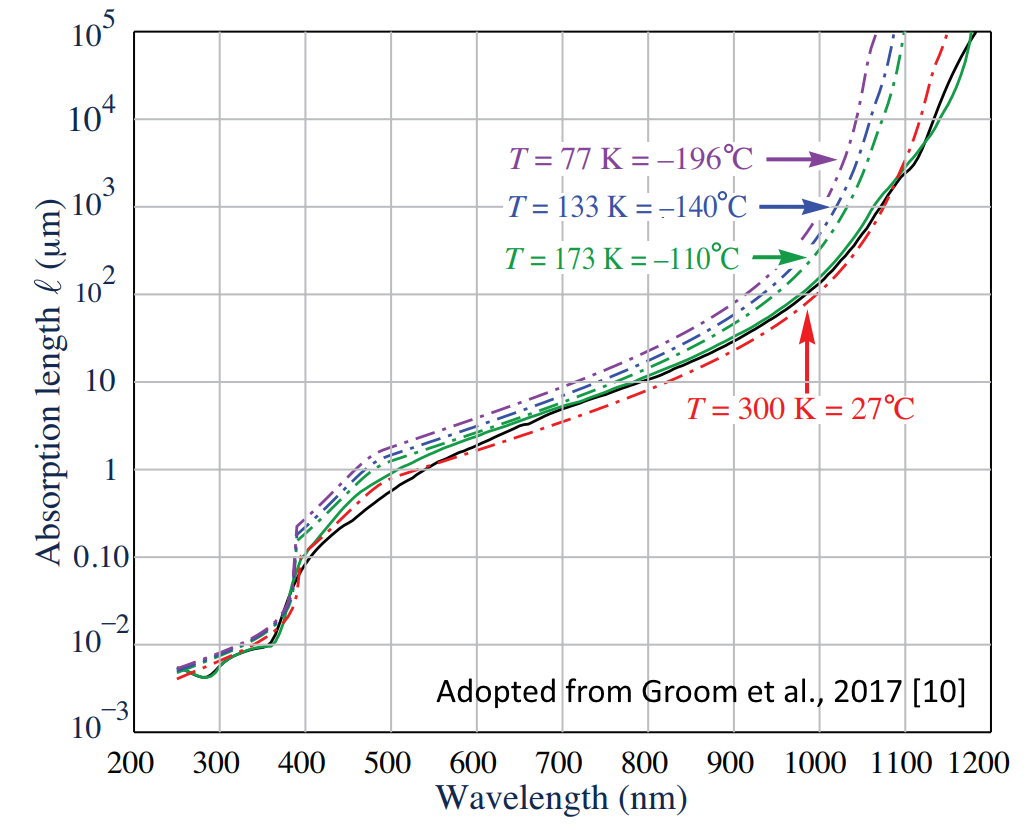}
    \caption{The absorption length in silicon, and therefore the IPRF, depends on both wavelength and temperature. Adopted from Figure 2 of Groom et al., 2017\cite{Groom2017}.}
    \label{fig:silicon_temp}
\end{figure}

Second, the effects of temperature on the absorption length in silicon (Figure \ref{fig:silicon_temp}) should be taken into account\cite{Groom2017}, before these maps can be implemented in the \textit{Kepler} data reduction tools. First, we will model the expected effect on the IPRF due to the difference between our operating temperature (-35$^{\circ}$ C) and \textit{Kepler}'s science operating temperature (-85$^{\circ}$ C). Next, we will acquire data at a range of temperatures (which are accessible with our camera) and compare them to the predicted equivalent wavelengths.

\section{Conclusions}
The spot scanning technique has been shown to be effective at measuring the IPRF of imaging sensors. Our initial data show that \textit{Kepler}'s IPRF is highly nonuniform and wavelength-dependent. Our new spot scanner uses modern technology (stable light sources, precise and reliable translation stages) to allow routine, non-destructive measurements of IPRF for a wide range of imaging sensors. 

\acknowledgments     
 
This research is funded by NASA grant NNX16AF43G. The authors would like to thank Charlie Sobeck, the Kepler Project Manager at NASA Ames, for assistance with locating and securing a Kepler CCD for these tests. We appreciate John Troeltzsch, Anne Ayers, and Rick Ortiz at Ball Aerospace in Boulder for providing information regarding the packaging and electronics associated with the Kepler CCD. We're grateful to Charles Slaughter and Gary Sims at Spectral Instruments in Arizona for modifying a camera to use the Kepler CCD. DV thanks J. Hoover for significantly improving the stability of the data acquisition software.


\bibliography{report}   
\bibliographystyle{spiebib}   

\end{document}